\documentclass{jltp}

\usepackage{graphicx} 

\title{Bound state of dimers on a spherical surface}

\author{M.K. Kostov, E.S. Hernandez$^*$ and M.W. Cole}

\address{Department of Physics, The Pennsylvania State University, University Park, PA 16802\\
$^*$Departamento de Fisica, Facultad de Ciencias Exactas y Naturales, Universidad de Buenos Aires,\\
and Consejo Nacional de Investigaciones Cient\'{\i}ficas y T\'ecnicas, 1428 Argentina}

\runninghead{M.K. Kostov, E.S. Hernandez and M.W. Cole}{Bound state of dimers on a spherical surface}

\begin{document}

\maketitle

\begin{abstract}
The study of particle motion on spherical surfaces is relevant to adsorption on buckyballs and other solid particles. This paper reports results for the binding energy of such dimers, consisting of two light particles (He atoms or hydrogen molecules) constrained to move on a spherical surface. The binding energy reaches a particularly large value when the radius of the sphere is about 3/4 of the particles' diameter.
\end{abstract}

\section{INTRODUCTION}
The problem of reduced dimensionality is one of the main themes in condensed matter physics. One reason for an explosion of interest in this field is the discovery of carbon nanotubes, which are quasi-one dimensional (1D). A similar growth of research activity has focused on clusters, which are zero-dimensional from the standpoint of statistical mechanics (because no length diverges). A typical problem in these fields is the relation between properties in the reduced dimension and those in our 3D world. For example, the critical temperature of the liquid-vapor transition T$_c$ increases monotonically from zero to a value of order the pair interaction energy as D increases from 1 to 3 or more. One might wonder, furthermore, about the evolution of behavior for systems of variable {\it effective} dimensionality D$_{eff}$ , such as the fluid adsorbed on a cylinder of varying radius. For very large (small) radius, D$_{eff}$ = 2 (1). 
 
In this paper, we evaluate the binding energy of two particles, whose nuclei are constrained to move on a spherical surface of radius R. This is related to recent work exploring the evolution of films adsorbed on a C$_{60}$ buckyball and similar substrates \cite{her,szy}. It is also related to an analogous study of dimer binding energy on a cylindrical surface \cite{kos}. That study found that dimers are very strongly bound if the cylinder's diameter is of order the hard-core diameter $\sigma$ of the interacting particles \cite{car}. In that situation, the binding energy greatly exceeds that found in the 2D limit of infinite radius. The reason for this large binding is the existence of a favorable phase space condition at a particular ratio of R/$\sigma$. We report here similar behavior in the spherical case; that is, a particularly large binding energy E$_B$ occurs if an analogous geometrical condition is satisfied. 
 
While this paper focuses on E$_B$ for the dimer, experience with the analogous cylindrical surface problem suggests that properties of the many-body system will also  exhibit extrema at a particular intermediate value of the ratio R/$\sigma$.\cite{kos,mer} 

\section{GROUND STATE ENERGIES AND WAVE FUNCTIONS}
One may think of this geometry by imagining that one particle resides at a fixed position, the pole, {\bf r}$_1$ =(x,y,z)=(0,0,R) while the other particle moves across the spherical surface in the presence of their interaction. The Schr$\ddot{o}$dinger equation uses the Hamiltonian:

\begin{equation}
H = \left[-\frac{{\hbar}^2}{2\mu}\right] \nabla_{2}^{2} + V(r) \, .
\end{equation}
Here $\mu$  is the interacting particles' reduced mass, $\nabla_{2}^{2}$  is the 2D Laplacian (variable polar and azimuthal angles, $\theta$ and $\phi$) and V(r) is taken as  the usual Lennard-Jones(LJ) interaction  with well-depth $\epsilon$. The coordinate r is the 3D separation: r=2R sin$\left(\theta/2\right)$. In all cases studied here, we have assumed the LJ parameters: for the He isotopes, $\sigma$=2.556 \AA\, and $\epsilon$ =10.22 K, while for H$_2$-H$_2$ the parameters are $\sigma$ = 3.04 \AA\, and $\epsilon$ = 34.3 K. Then, Eq. (1) can be reduced to a dimensionless form; the resulting equation provides an implicit relation between the reduced binding energy E$_{B}^*$ =E$_B$/$\epsilon$ and the ratio R/$\sigma$. This reduced equation involves a dimensionless quantity $\displaystyle{h / \left[\sigma (\mu\epsilon)^{1/2}\right]}$, related to the de Boer quantum parameter, discussed below; when this quantity is large, the reduced zero-point energy is large and the binding energy is small. 
 
A useful function is the specific area $a(r)$, defined as the area of the sphere lying within a distance interval [r, r+dr] from  {\bf r}$_1$, divided by dr. A simple calculation yields the following result: $a(r) = 2 \pi r \Theta(2R-r)$, where $\Theta(x)$ is the Heaviside unit step function. This result means that the most surface area lies at distance from {\bf r}$_1$ nearly equal to 2R, but there exists no surface at separation r$>$2R. To take optimal advantage of the attraction, therefore, the potential minimum should lie at distance somewhat less than 2R. One can prove the following consequence of this geometry: the classical second virial coefficient of a fluid on this surface B$_{class}$(T) has a maximum when R=$\sigma$/2 and decreases monotonically with R if R$>\sigma$/2. One might investigate whether similar behavior occurs in the quantum case. Then, the low temperature behavior is dominated by the spectrum of bound states, if any are present. 

\begin{figure}
\includegraphics[height=3.00in]{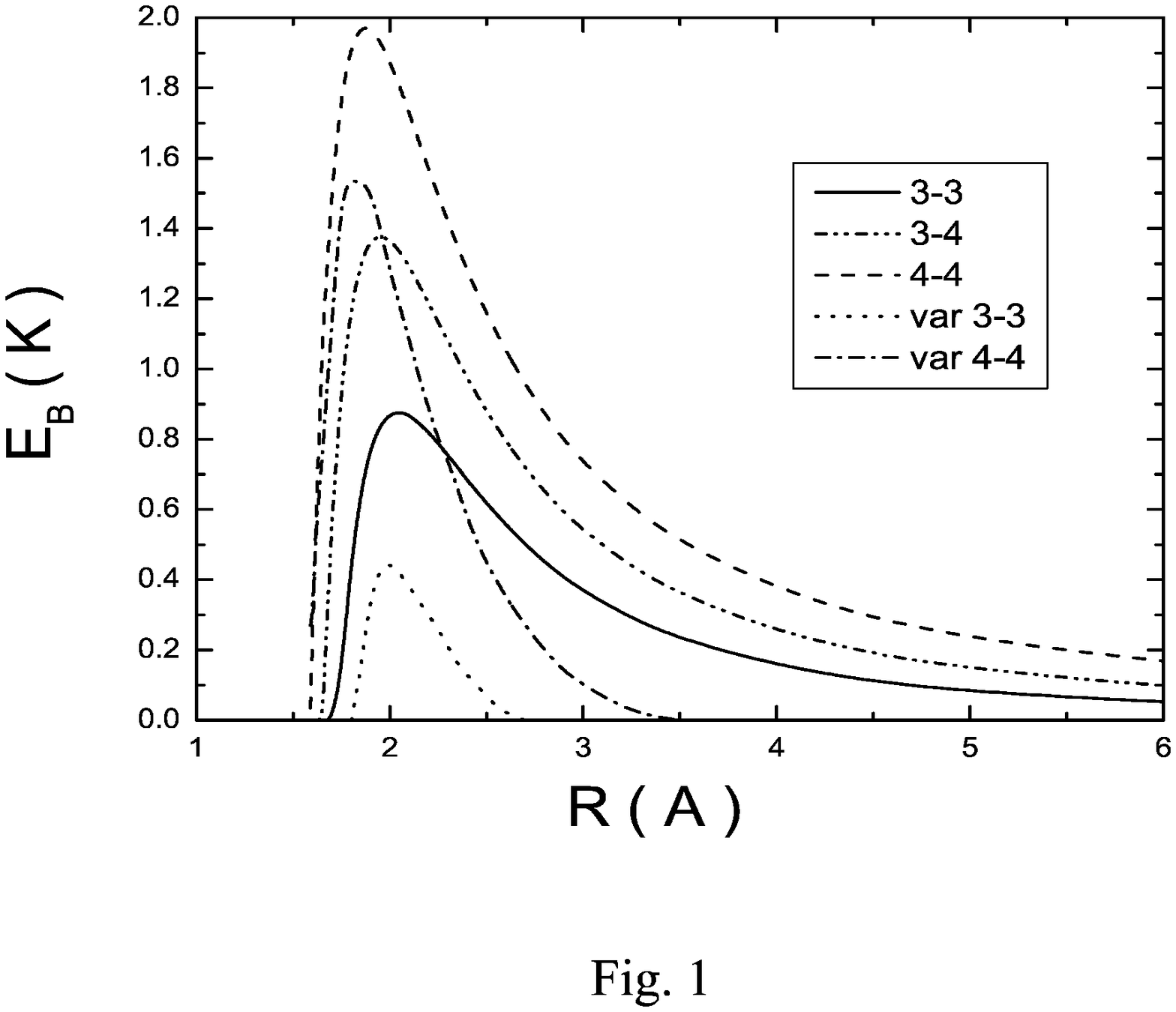}
\caption{Binding energies for the various He dimer problems as a function of sphere radius. The curves labeled ``var'' are variational results while the others are exact.}  
\label{fig1}
\end{figure}
 
We have determined the ground state energy (-E$_B$) for this problem in two ways, variational and exact, for a number of different systems, including $^3$He -$^3$He, $^3$He -$^4$He and $^4$He -$^4$He, as well as H$_2$-H$_2$. An interesting question is this: what ratio of R/$\sigma$ gives the largest binding energy E$_B$. Since no attraction is provided by the LJ potential unless R/$\sigma>$1/2, one might guess that the largest binding occurs when R/$\sigma \sim$1. Furthermore, we expect the system to attain the largest reduced energy E$_{B}^*$  when the particles are most classical; a completely classical system would have E$_{B}^*$ =1 since the particles would maintain a spacing equal to the minimum in the potential. The criterion for ``most classical'' is the largest value of the de Boer quantum parameter: $\displaystyle{\Lambda^{*} = \frac{h}{\left[\sigma (m \epsilon)^{1/2}\right]}}$ where $m$ is the mass of the interacting particles, assumed to be the same. The values are $\Lambda^*$= 3.1, 2.7 and 1.7 for the sequence $^3$He, $^4$He and H$_2$, respectively .

Figure 1 presents the exact and variational energies for the various He cases. There is seen to be a maximum binding energy of order 1 to 2 K, with specific values 1.97 K, 1.38 K and 0.88 K for the sequence $^4$He - $^4$He, $^3$He - $^4$He and $^3$He - $^3$He, respectively. This trend manifests the expected decrease of binding energy with increasing $\Lambda^{*}$; a similar behavior was reported by Kili$\acute{c}$ {\it et al.} for dimers confined to either flat surfaces or to the interior of an spherical cavity \cite{kil1,kil2}. The maximum binding occurs with  R/$\sigma \sim$ 0.72, 0.76 and 0.80. This is analogous to the case of dimer binding of He isotopes on a cylindrical surface\cite{kos}, with a (somewhat smaller) maximum binding at R/$\sigma$ =0.65. The smaller value of R/$\sigma$ in the cylindrical case is expected because the a(r) cutoff of long range contributions to the attraction, mentioned above, is not present there.  

\begin{figure}
\includegraphics[height=3.00in]{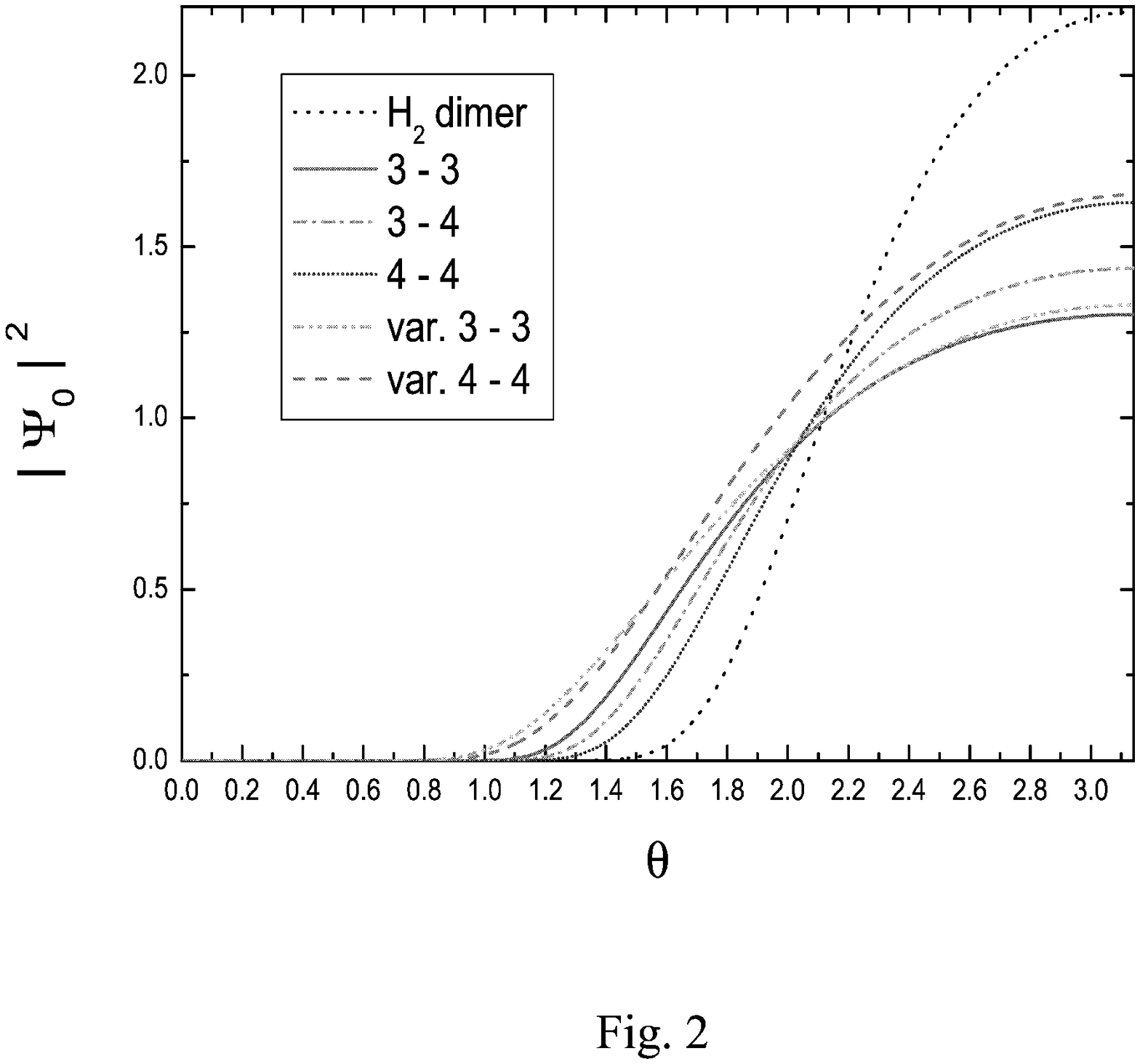}
\caption{Normalized probability density as a function of polar angle for the various cases studied here.}  
\label{fig2}
\end{figure}
 
Figure 1 also shows the result of a variational calculation, using a trial wave function $\Psi_{v}(r)$ of the form:

\begin{equation} 
 \Psi_{v}(r) = \exp \left[-\frac{1}{2}\left(\frac{b \sigma}{r} \right)^{n}\, \right] \,.
\end{equation} 
 The two variational parameters b and n were optimized for each radius. The resulting energy in the $^3$He - $^3$He case, shown in Fig. 1, has  b=1.80 and n=2.50 at the optimal radius R/$\sigma$=0.80. Note that the variational result for the binding energy is significantly lower than the exact binding energy, while the optimized radius  is nearly identical to the exact one, R/$\sigma$ =0.80. An analogous situation was found in the problem of the dimer on a cylinder\cite{kro}. There, the optimized variational binding energy was 0.084 K, while the exact energy  was 0.218 K. Interestingly, the two calculations yielded the same optimal radius in that case too. 
 
 Figure 2 compares the probability densities for the various cases studied here. The characteristic feature is an increasing localization near the south pole as the system becomes more classical, as implied by the $a(r)$ discussion above. Note that the probabilities derived from the variational wave functions are more diffuse than the exact solutions.
 
\begin{figure}
\includegraphics[height=3.00in]{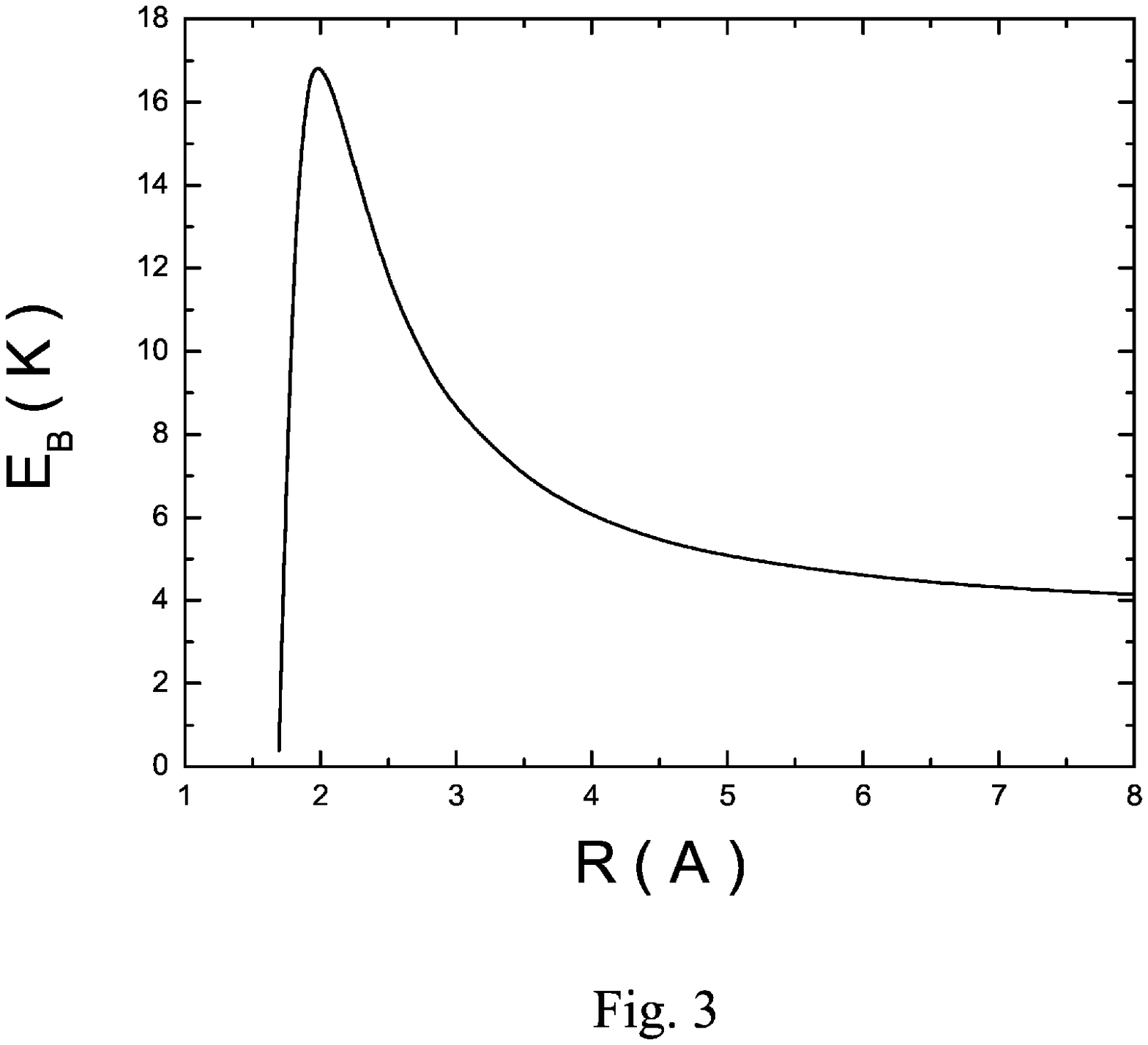}
\caption{Binding energy for H$_2$ - H$_2$ as a function of sphere radius.}  
\label{fig3}
\end{figure}

In Fig. 3 we present the exact binding energy for the H$_2$ - H$_2$ dimer. In this case, the most classical of the group reported here, the maximum  E$_{B}^{*}$ is 0.5. This value is much closer to the classical value (1) than that for  $^4$He - $^4$He (E$_{B}^{*}$=0.2). One might not have anticipated such a large increase in E$_{B}^{*}$ since the   $\Lambda^*$ parameters are not so different ( 2.7 vs 1.8). Note, however, that for the He isotopes, just a 15$\%$ difference between values of $\Lambda^*$ leads to a factor of two difference in  E$_{B}^{*}$. Evidently, the quantum parameter plays a critical role in this problem because the hard core potential confines the atoms to a fairly small domain on the surface.

\section{SUMMARY}
We have found particularly large binding energies for dimers of He and H$_2$ when the particles are constrained to move on spherical surfaces having R/$\sigma \sim $3/4. This behavior is qualitatively similar to that found previously for cylindrical surfaces. A possible next step would be to explore the quantum virial expansion on this surface. Such a theory would take a different form from the phase shift approach used for extended systems. From the experimental point of view, the present work suggests the intriguing possibility of spectroscopic observations of these dimer states, e.g. in a mixture of C$_{60}$ molecules and He atoms. A quantitative prediction would require a study of both the excited state spectrum and the role of the third dimension, associated with the radial motion, which is ignored in the present work.

\section*{ACKNOWLEDGMENTS}
This research is supported by NSF grants DMR-0208520 and INT-0124087, SANCOM collaboration between Penn State and Air Products and Chemicals, Inc., ANPCYT($\#$PICT03-08450) and University of Buenos Aires ($\#$UBA-X103).



%
%

\end{document}